\documentstyle[prl,aps,epsfig,multicol]{revtex}

\begin{document}


\draft

\title{Lifetimes of agents under external stress}

\author{Claus O. Wilke and Thomas Martinetz}

\address{Institut f\"ur Neuroinformatik\\Ruhr-Universit\"at Bochum}

\date{Submitted: December 8, 1998; Printed: \today}

\maketitle

\begin{abstract}
An exact formula for the distribution of lifetimes in
coherent-noise models and related models is derived. For certain
stress distributions, this formula can be analytically evaluated
and yields simple closed expressions. For those types of stress for
which a closed expression is not available, a numerical evaluation
can be done in a straightforward way. All results obtained are
in perfect agreement with numerical experiments. The implications for the
coherent-noise models' application to macroevolution are discussed.
\end{abstract}

\pacs{PACS numbers: 
05.40.-a, 
87.23.Kg 
}


\begin{multicols}{2}
Agents under externally imposed stress have been recently studied in
coherent-noise and related models~\cite{NewmanSneppen96,Newman96,%
SneppenNewman97,WilkeMartinetz97,WilkeAltmeyerMartinetz98a,%
WilkeMartinetz98,Standish99}. These models display scale free
distributions in a number of 
quantities, such as event sizes and lifetimes, or in the decay pattern
of aftershocks. Coherent-noise models are very different from other
models displaying scale free behavior, such as sand pile
models~\cite{BakTangWiesenfeld87}, 
as they do not rely on local interactions or feedback.  Hence,
they are not self organized critical. Considered the abundance of
power-law distributed quantities in nature~\cite{Bak97}, models such
as the ones of the coherent-noise type can help understanding to what
extent self-organized criticality is the right paradigm for describing
driven systems, and to what extent other mechanisms can provoke
similar power-law distributions.

Despite the simplicity of the original coherent-noise model---agents
have thresholds $x_i$; if global stress exceeds a threshold $x_i$,
agent $i$ gets replaced; with prob. $f$, an agent gets a new
threshold---, no exact analytical results have been obtained so
far. The distributions of event sizes and aftershocks have been
studied in detail in~\cite{SneppenNewman97} (event sizes) and
in~\cite{WilkeAltmeyerMartinetz98} (aftershocks). Both distributions
can be regarded as being well understood. Nevertheless, the
theoretical results are only of approximative character in both cases.

In the case of the distribution of lifetimes, there are even less
theoretical results. Sneppen and Newman~\cite{SneppenNewman97} have
given an expression based 
on their time-averaged approximation. This expression is right for
certain stress distributions, as we will show below. However, it
breaks down for slowly decaying distributions such as the Lorentzian
distribution. Moreover, it is not clear when exactly it can be
applied. 

In a recent paper~\cite{Standish99}, a different approach of
calculating the distribution of lifetimes has been taken, and
the author claimed that the lifetimes obey multiscaling, with a
$L^{-2}$ decrease for small lifetimes, and a $L^{-1}$ decrease for
large lifetimes. Here, we will demonstrate that this statement is
wrong. We will calculate the distribution of lifetimes exactly,
without any approximations, and we will show that our results are in
perfect agreement with numerical simulations. 

Our calculations are based on the observation that it is not necessary
to know the distribution of thresholds $\rho(x)$ for calculating the
distribution of lifetimes. All we have to know is the distribution according to
which agents enter the system, which is called $p_{\rm thresh}(x)$ in
the notation of~\cite{NewmanSneppen96}, and the stress distribution
$p_{\rm stress}(x)$. Once an agent has entered the
system, it has a well defined life expectancy, which is closely
related to the probability that the agent will be hit by stress or
mutation. Note that in this picture, we are considering only a single
agent. Therefore, if we talk about lifetimes, it does not matter
whether the stress acts coherently on a large number of agents, or
whether it is drawn for all agents independently. In this respect, the
results we obtain in this work are of a much more general nature than
the results found previously for event sizes or aftershocks.

An agent with threshold $x$ will survive stress and mutation in one
time step with a
probability $p(x)$ equal to~\cite{WilkeAltmeyerMartinetz98a}
\begin{eqnarray}\label{eq:def_p_of_x}
 p(x) &=& (1-f)[1-p_{\rm move}(x)]\nonumber\\
      &=& (1-f)\int_0^x p_{\rm stress}(x')\, dx'\,.
\end{eqnarray}
What is the distribution of the survival probabilities $p$? We 
denote the corresponding density function by $u(p)$. Clearly, we have
\begin{eqnarray}
  u(p)\,dp &=& p_{\rm thresh} (x) \, dx \nonumber\\
    &=& dx\quad \mbox{for}\quad 0\leq x <1\,.
\end{eqnarray}
In the second step, we have assumed that the threshold distribution is
uniform. This can always be achieved after a suitable transformation
of variables~\cite{SneppenNewman97}. Hence, we find
\begin{equation}
  u(p)=\frac{dx}{dp}\,.
\end{equation}
The derivative $dx/dp$ can be calculated from
Eq.~(\ref{eq:def_p_of_x}),
\begin{equation}\label{eq:dx_by_dp}
\frac{dx}{dp}=\frac{1}{(1-f)p_{\rm stress}[x(p)]}\,,
\end{equation}
and $x(p)$ can be obtained from Eq.~(\ref{eq:def_p_of_x}) by
inversion. The density function is thus defined for $p< p_{\max}$, where
\begin{equation}\label{eq:def_p_max}
  p_{\max} = p(1) = (1-f)\int_0^1 p_{\rm stress}(x)\, dx
\end{equation}
stems from the condition that the thresholds are distributed
uniformly between 0 and 1. Above $p_{\max}$, the density function
$u(p)$ is equal to zero. 

All agents with the same survival probability $p$ generate a
distribution of lifetimes which reads
\begin{equation}\label{eq:def_g_of_L}
  g(L) = p^{L-1}(1-p)\,.
\end{equation}
Here, $g(L)$ is the probability density function for the lifetimes
$L$. Note that the model works in discrete time steps, therefore the
lifetimes $L$ are integers, and $g(L)$ is only defined for integral
arguments.
We can calculate the distribution of lifetimes $h(L)$ by averaging
over the distributions generated by different 
survival probabilities $p$, weighted with their density function
$u(p)$:
\begin{equation}\label{eq:def_h_of_L}
  h(L)=\int_0^{p_{\max}} u(p)p^{L-1}(1-p)\,dp\,.
\end{equation}

Equation~(\ref{eq:def_h_of_L}) can be explicitly evaluated for
exponentially distributed stress, $p_{\rm stress}=
\exp(-x/\sigma)/\sigma$. We find 
\begin{equation}
  u(p)=\frac{\sigma}{1-f-p}\quad \mbox{for}\quad 0\leq p <p_{\max}\,,
\end{equation}
with 
\begin{equation}
  p_{\max}=(1-f)[1-\exp(-1/\sigma)]\,.
\end{equation}
After inserting this into Eq.~(\ref{eq:def_h_of_L}) and doing some
basic calculations, we obtain
\begin{equation}\label{eq:h_of_l_exp}
  h(L)=\frac{\sigma p_{\max}^L}{L} 
       + f\sigma\int_0^{p_{\max}}\frac{p^{L-1}}{1-f-p}\,dp\,.
\end{equation}
 It is possible to calculate the remaining
integral with the aid of the identity (see~\cite{AbramowitzStegun84},
15.3.1)
\begin{eqnarray}
&&\int_0^1 t^{b-1}(1-t)^{c-b-1}(1-tz)^{-a}\,dt= \nonumber\\
&& \qquad\qquad\qquad= \frac{\Gamma(b)\Gamma(c-b)}{\Gamma(c)}F(a,b;c;z)\,,
\end{eqnarray}
where $F(a,b;c;z)$ is the hypergeometric function. We find
\begin{equation}\label{eq:h_of_L_exp}
h(L)=\sigma\frac{p_{\max}^L}{L}\left[1+\frac{f}{1-f}
        F\left(L,1;L+1;\frac{p_{\max}}{1-f}\right)\right]\,.
\end{equation}
The leading term $\sigma p_{\max}^L/L$ is responsible for a $L^{-1}$
decay with cut off at $L \approx 1/f$. This behavior has been reported
previously, and it corresponds to the approximation derived
in~\cite{SneppenNewman97}. The correcting term vanishes 
with $f$. It is of importance only for extremely long lifetimes of
the order $1/f$, for which it modifies the detailed cut off behavior.

\begin{figure}
\narrowtext
\centerline{ 
 \epsfig{file={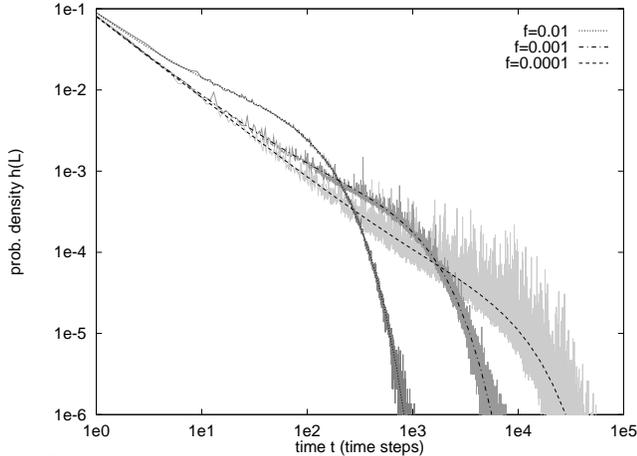}, width=\columnwidth}}
\caption{The distribution of lifetimes of agents subjected to  exponentially
  distributed stress, with $\sigma=0.08$. The gray lines represent
  the results obtained from computer experiments, the black lines
  represent the theoretical prediction
  Eq.~(\ref{eq:h_of_l_exp}). Theory and computer experiment are in perfect agreement.
\label{fig:eps-stress}}
\end{figure}

In Fig.~\ref{fig:eps-stress} we display Eq.~(\ref{eq:h_of_L_exp}) together with
results from direct numerical simulations, for different values of
$f$. The theoretical result is in perfect agreement with the measured
distributions. The dependency of the cut off on $f$ is clearly visible
in Fig.~\ref{fig:eps-stress}.

\begin{figure}
\narrowtext
\centerline{ 
 \epsfig{file={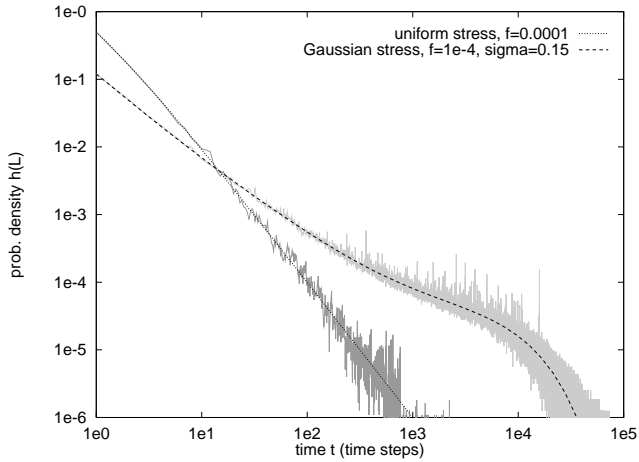}, width=\columnwidth}}
\caption{Comparison of theory and computer experiment for the uniform
  and for the Gaussian stress distribution. As in the case of
  Fig.~\ref{fig:eps-stress}, we observe perfect agreement.
\label{fig:uniform-gauss}}
\end{figure}

Another stress distribution for which we can derive a closed analytic
form for $h(L)$ is the uniform distribution, $p_{\rm stress}(x)=1$ for
$0\leq x <1$. We find
\begin{equation}
  u(p) = \frac{1}{1-f}\quad \mbox{for}\quad 0\leq p<1-f
\end{equation}
and
\begin{equation}\label{eq:h_of_l_uniform}
  h(L) = \frac{(1-f)^{L-1}}{L(L+1)}(1+fL)\,.
\end{equation}
As in the case of Eq.~(\ref{eq:h_of_l_exp}), we get a leading and a
correcting term. The leading term decays as $L^{-2}$ with cut-off at
$L\approx 1/f$, and the second term modifies the cut-off
behavior. Interestingly, the distribution of lifetimes is scale-free,
although the distribution of event sizes in a coherent-noise model
with uniform stresses is not a power law~\cite{SneppenNewman97}.
A plot of Eq.~(\ref{eq:h_of_l_uniform}) is given in
Fig.~\ref{fig:uniform-gauss}, together with the corresponding
measured distribution. 

\begin{figure}
\narrowtext
\centerline{ 
 \epsfig{file={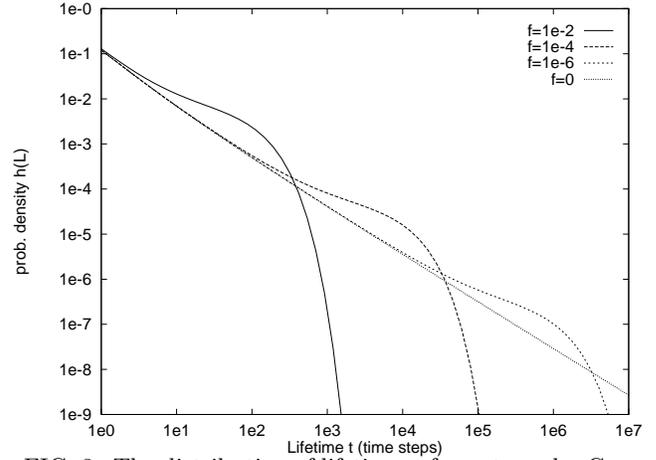}, width=\columnwidth}}
\caption{The distribution of lifetimes of agents under Gaussian
  stress, with $\sigma=0.15$ and different values of $f$. For very
  small mutation rates $f$, the distribution of lifetimes becomes
  $1/L$ for $L$ larger than about $10^5$ time steps.
\label{fig:gaussian}}
\end{figure}

For the most other stress distributions, the integral in
Eq.~(\ref{eq:def_h_of_L}) can only be done numerically. This is the
case, for example, for the Gaussian distribution, $p_{\rm
  stress}(x)=\sqrt{2/(\pi\sigma^2)}\exp[-x^2/(2\sigma^2)]$. 
Under Gaussian stress, an agent with threshold $x$ will survive a
single time step with probability
\begin{equation}\label{eq:p_of_x_gauss}
  p(x)=(1-f)\, {\rm erf}\left(\frac{x}{\sqrt{2}\sigma}\right)\,,
\end{equation}
where ${\rm erf}(x)$ is the error function
\begin{equation}
{\rm erf}(x)=\frac{2}{\sqrt{\pi}}\int_0^x \exp(-t^2)\,dt\,.
\end{equation}
Inversion of Eq.~(\ref{eq:p_of_x_gauss}) yields
\begin{equation}\label{eq:x_of_p_gauss}
  x(p) = \sqrt{2}\sigma\, {\rm erf}^{-1}\left(\frac{p}{1-f}\right)\,.
\end{equation}
Here, by ${\rm erf}^{-1}(z)$ we denote the inverse error function,
obtained by solving the equation $z={\rm erf}(x)$ for $x$.
We can calculate the density function of the survival probabilities
with the aid of Eqs.~(\ref{eq:dx_by_dp}) and~(\ref{eq:x_of_p_gauss}). The
resulting expression reads
\begin{equation}
  u(p)=\sqrt{\frac{\pi}{2}}\frac{\sigma}{1-f} \exp\left( 
    \left[{\rm erf}^{-1} \left(\frac{p}{1-f}\right)\right]^2\right)\,.
\end{equation}
The numerical integration of $u(p)p^{L-1}(1-p)$ is somewhat tricky for
choices of $\sigma$ such, that $p_{\max}/(1-f)$ is very close to~1,
since the inverse error function has a singularity at~1. However, for
moderately small $\sigma$, the integration can be carried out without
too much trouble. The resulting density function $h(L)$ is shown in
Fig.~\ref{fig:uniform-gauss} for $\sigma=0.15$ and $f=10^{-4}$. We
find that, for $L\ll1/f$, the function $h(L)$ is almost linear in the
log-log plot. A fit to the linear region of $h(L)$ gives an exponent
$\tau=1.177\pm 0.01$, which means $h(L)$ decays slightly steeper than
the $L^{-1}$ decay predicted by the approximation of Sneppen and Newman. 
However, if we evaluate $h(L)$ for much larger $L$ and much smaller
$f$, we find that the exponent $\tau$ decreases slowly towards the
value 1 (Fig.~\ref{fig:gaussian}).

\begin{figure}
\narrowtext
\centerline{ 
 \epsfig{file={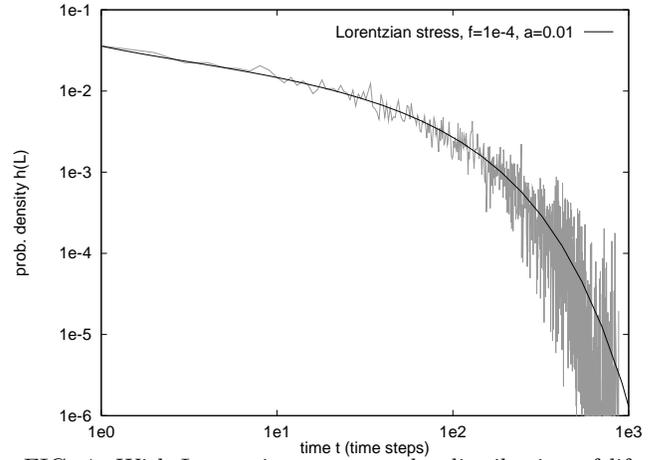}, width=\columnwidth}}
\caption{With Lorentzian stresses, the distribution of
  lifetimes is no longer scale invariant.
\label{fig:lorentzian}}
\end{figure}

Let us now turn to the Lorentzian distribution $p_{\rm stress}(x)
=(2a/\pi)/(x^2+a^2)$. In this case, a calculation along the lines
of Eqs.~(\ref{eq:def_p_of_x})--(\ref{eq:def_p_max}) yields the following
distribution of survival probabilities:
\begin{equation}
  u(p)=\frac{\pi}{2}\frac{a}{1-f}\left(\cos^2
       \left[\frac{\pi}{2}\frac{p}{1-f}\right]\right)^{-1}\,.
\end{equation}
Here, $p_{\rm max}=(2/\pi)(1-f)\arctan (1/a)$. The result of the
numerical integration is shown in Fig.~\ref{fig:lorentzian}. As in
the previous cases, we observe a perfect agreement between the
analytic expression for $h(L)$ and the distribution measured in
computer experiments. In the case of Lorentzian stresses, the
distribution of lifetimes is clearly not scale invariant.

In~\cite{Standish99} it has been claimed that the distribution of the
agents' lifetimes under external stress decays as $L^{-2}$ for small
$L$. Among the four stress distributions considered in this work, we
found a $L^{-2}$ decay only for the uniform stress distribution. Hence
the statement made in~\cite{Standish99} is wrong in general. We could
verify the $L^{-1}$ decay reported in~\cite{SneppenNewman97} for
exponential or Gaussian stresses. As it was also stated there, the
Lorentzian stress distribution does not produce a scale free
distribution of lifetimes.

A surprising result of this work is the observation that the
properties of the distribution of lifetimes and of the distribution of
event sizes in a coherent-noise model are largely independent from
each other. We do find power-law distributed lifetimes under uniform
stress, under which the distribution of event sizes is not scale free,
and we do not find power-law distributed lifetimes under Lorentzian
stress, which generates a scale free distribution of event
sizes. Consequently, we cannot infer from a power-law distribution of
event sizes to one of lifetimes, and vice versa. Both distributions
have to be investigated independently for every type of stress.

Let us conclude with some remarks on the implications of our results
for the application of coherent-noise or related models to large scale
evolution. In the context of macroevolution, the agents are regarded
as species, or higher taxonomical units, such as genera or
families~\cite{Newman98}. The 
distribution of genus lifetimes in the fossil record follows either
a power-law decrease with exponent near 2, or an exponential
decrease~\cite{Sneppenetal95,NewmanSibani99}. A $L^{-2}$ decay can be
observed in coherent-noise models with uniform stress. However, in
this case the distribution of extinction events does not follow the
$s^{-2}$ decay -- with $s$
denoting the number of families gone extinct in one time step -- found
in the fossil record~\cite{Newman96}. 
The distribution of lifetimes closest to an exponential decay is,
among the stress distributions we studied here,   
generated by Lorentzian stresses. But also in this case, the
distribution of extinction events is significantly different from the
needed $s^{-2}$ decay of extinction events. On the other hand, it
seems to be typical for 
distributions generating a $s^{-2}$ decay, such as exponential,
Gaussian, or Poissonian, that the distribution of lifetimes decays as
$L^{-1}$. It is arguable whether any type of stress can actually
generate the right type of distribution for lifetimes and extinction
events simultaneously. Hence, the coherent-noise models in their
current formulation probably miss some important ingredient as a model
of macroevolution. An effect
which is not covered, and which has been shown recently to be of
importance for the statistical patterns in the fossil record, is a
decline in the extinction rate~\cite{NewmanSibani99,RaupSepkoski82,SibaniSchmidtAlstrom95,NewmanEble99}. For example, Sibani 
\emph{et al.}~\cite{SibaniSchmidtAlstrom95,Sibani97} have
demonstrated that the $L^{-2}$ decay in lifetimes might be closely
related to the decline in the extinction rate.


\end{multicols}

\end{document}